\documentclass[twoside,twocolumn,9pt]{article}
\usepackage{extsizes}
\usepackage[numbers]{natbib}
\usepackage[version=3]{mhchem}
\usepackage[left=1.5cm, right=1.5cm, top=1.785cm, bottom=2.0cm]{geometry}
\usepackage{balance}
\usepackage{times,mathptmx}
\usepackage{sectsty}
\usepackage{graphicx} 
\usepackage{lastpage}
\usepackage[format=plain,justification=justified,singlelinecheck=false,font={stretch=1.125,small,sf},labelfont=bf,labelsep=space]{caption}
\usepackage{float}
\usepackage{fancyhdr}
\usepackage{fnpos}
\usepackage[english]{babel}
\addto{\captionsenglish}{%

}
\usepackage{array}
\usepackage{droidsans}
\usepackage{charter}
\usepackage[T1]{fontenc}
\usepackage[usenames,dvipsnames]{xcolor}
\usepackage{setspace}
\usepackage[compact]{titlesec}
\usepackage{hyperref}
\usepackage{afterpage}

\usepackage{epstopdf}

\definecolor{cream}{RGB}{222,217,201}

\begin{document}

\pagestyle{fancy}
\thispagestyle{plain}

\makeFNbottom
\makeatletter
\renewcommand\LARGE{\@setfontsize\LARGE{15pt}{17}}
\renewcommand\Large{\@setfontsize\Large{12pt}{14}}
\renewcommand\large{\@setfontsize\large{10pt}{12}}
\renewcommand\footnotesize{\@setfontsize\footnotesize{7pt}{10}}
\makeatother

\renewcommand{\thefootnote}{\fnsymbol{footnote}}
\renewcommand\footnoterule{\vspace*{1pt}%
\color{cream}\hrule width 3.5in height 0.4pt \color{black}\vspace*{5pt}} 
\setcounter{secnumdepth}{5}

\makeatletter 
\renewcommand\@biblabel[1]{#1}            
\renewcommand\@makefntext[1]%
{\noindent\makebox[0pt][r]{\@thefnmark\,}#1}
\makeatother 
\renewcommand{\figurename}{\small{Fig.}~}
\sectionfont{\sffamily\Large}
\subsectionfont{\normalsize}
\subsubsectionfont{\bf}
\setstretch{1.125} 
\setlength{\skip\footins}{0.8cm}
\setlength{\footnotesep}{0.25cm}
\setlength{\jot}{10pt}
\titlespacing*{\section}{0pt}{4pt}{4pt}
\titlespacing*{\subsection}{0pt}{15pt}{1pt}

\fancyfoot{}
\fancyfoot[LO]{submitted to PCCP}
\fancyfoot[RE]{uploaded to arxiv}
\fancyfoot[RO]{\footnotesize{\thepage}}
\fancyfoot[LE]{\footnotesize{\thepage}}
\fancyhead{}
\renewcommand{\headrulewidth}{0pt} 
\renewcommand{\footrulewidth}{0pt}
\setlength{\arrayrulewidth}{1pt}
\setlength{\columnsep}{6.5mm}
\setlength\bibsep{1pt}

\makeatletter 
\newlength{\figrulesep} 
\setlength{\figrulesep}{0.5\textfloatsep} 

\newcommand{\topfigrule}{\vspace*{-1pt}%
\noindent{\color{cream}\rule[-\figrulesep]{\columnwidth}{1.5pt}} }

\newcommand{\botfigrule}{\vspace*{-2pt}%
\noindent{\color{cream}\rule[\figrulesep]{\columnwidth}{1.5pt}} }

\newcommand{\dblfigrule}{\vspace*{-1pt}%
\noindent{\color{cream}\rule[-\figrulesep]{\textwidth}{1.5pt}} }

\makeatother
\twocolumn[
  \begin{@twocolumnfalse}
\sffamily
\begin{tabular}{m{0cm} p{18cm} }

 & \noindent\LARGE{\textbf{Rectification of bipolar nanopores in multivalent electrolytes: effect of charge inversion and strong ionic correlations}} \\
\vspace{0.3cm} & \vspace{0.3cm} \\

 & \noindent\large{D\'avid Fertig,$^{\ast}$\textit{$^{a}$} M\'onika Valisk\'o,\textit{$^{a}$} and Dezs\H{o} Boda\textit{$^{a}$}} \\

 & \noindent\normalsize{
Bipolar nanopores have powerful rectification properties due to the asymmetry in the charge pattern on the wall of the nanopore.
In particular, bipolar nanopores have positive and negative surface charges along the pore axis.
Rectification is strong if the radius of the nanopore is small compared to the screening length of the electrolyte so that both cations and anions have depletion zones in the respective regions.
The depths of these depletion zones is sensitive to sign of the external voltage.
In this work, we are interested in the effect of the presence of strong ionic correlations (both between ions and between ions and surface charge) due to the presence of multivalent ions and large surface charges.
We show that strong ionic correlations cause leakage of the coions, a phenomenon that is absent in mean field theories.
In this modeling study, we use both the mean-field Poisson-Nernst-Planck (PNP) theory and a particle simulation method, Local Equilibrium Monte Carlo (LEMC), to show that phenomena such as overcharging and charge inversion cannot be reproduced with PNP, while LEMC is able to produce nonmonotonic dependence of currents and rectification as  a function of surface charge strength.

} \\

\end{tabular}

 \end{@twocolumnfalse} \vspace{0.6cm}

  ]

\renewcommand*\rmdefault{bch}\normalfont\upshape
\rmfamily
 \section*{}
 \vspace{-1cm}


\footnotetext{\textit{$^{a}$~Department of Physical Chemistry, University of Pannonia, P. O. Box 158, H-8201 Veszpr\'em, Hungary}}






\section{Introduction}
\label{sec:intro}

Strong ionic correlations cause peculiar behavior in devices containing electrolytes.
The strength of the correlation can be increased by decreasing the temperature,~\cite{henderson_cpl_2000} changing the solvent to something that screens more weakly (organic solvents~\cite{parsons_ea_1976}), or getting rid of it altogether (molten salts.~\cite{boda_jcp_1999,lanning_jpcb_2004} and ionic liquids~\cite{vatamanu_jpcm_2016})
One can measure phenomena caused by strong ionic correlations in room-temperature aqueous electrolytes (most commonly used in nanofluidic devices) if the electrolyte contains multivalent ions.~\cite{loessberg_zahl_ac_2016,chou_nl_2018,ramirez_jmembsci_2018,wang_jpcc_2018,nasir_jcis_2019,li_jpcc_2019,besteman_prl_2004}

The most common example is observed when the electrolyte is near a charged wall (a metal electrode or charged groups of an insulator, for example), and the multivalent ions overcharge the wall because they correlate strongly with the surface charge.~\cite{valisko_aipadv_2018,devos_acis_2019,voukadinova_jcp_2019}
Overcharging simply means that more counterions are attracted to the surface than necessary to compensate the surface charge. 
In the resulting electrical double layer, a layer of excess coions appears that produces a change in the sign of the electrical potential in this layer, a phenomenon known as charge inversion.

In this work, we focus on overcharging caused purely by electrostatic interactions.
This phenomenon was heuristically explained by Shklovskii's theory on strongly correlated liquids near a charged surface.~\cite{shklovskii_prl_1999,nguyen_prl_2000,nguyen_nato_2001,grosberg_revmodphys_2002}
The basic idea of the mechanism is that multivalent counterions can overcharge the interface if their average distance is larger than the screening length so they ``do not feel each other''.
Computer simulations validated this theory because they naturally sample all the configurations that are important for strong correlations.

Other mechanisms of overcharging are based on interactions that are not screened electrostatic interactions.
A short range chemical interaction can cause specific chemical adsorption~\cite{lidonlopez_ecc_2014}, for example.
If small counterions and large coions are present, steric forces can induce overcharging.~\cite{greberg_jcp_1998}
These effects are absent in the model used in this study.

Charge inversion causes many experimental phenomena such as the attraction between like-charged particles,~\cite{lyubartsev_prl_1998,allahyarov_jpcm_2005} the reversal of the sign of the electrophoretic mobility~\cite{jimenez_l_2012}, or the reversal of the selectivity of nanopores.~\cite{he_jacs_2009,garciagimenez_pre_2010,garcia_gimenez_bri_2012,ramirez_jmembsci_2018,nasir_jcis_2019}
In the latter example, a negatively charged nanopore that is cation selective for KCl becomes anion selective for a 3:1 electrolyte.~\cite{he_jacs_2009}
In that system, the principal device function is selectivity.
Here, as in our previous studies,~\cite{hato_pccp_2017,madai_jcp_2017,matejczyk_jcp_2017,madai_pccp_2018,valisko_jcp_2019,madai_jml_2019,fertig_mp_2018,madai_pccp_2019,fertig_jpcc_2019} by device function we mean (often dimensionless) quantities that are put together from output signals of the device, which are ionic currents in the case of nanopores.
Selectivity, for example, can be defined as $I_{i}/I$, where $I_{i}$ is the ionic current carried by ionic species $i$ and $I$ is the total current.

Our study is about a nanopore that carries a bipolar surface charge pattern on its wall.
These kinds of nanopores have a region that carries positive (``p'' region) and an adjacent region that carries negative (``n'' region) surface charge.
These regions have different selectivity and conduction properties that together result in the overall selectivity and conduction properties of the whole pore.
Furthermore, the bipolar nanopore is asymmetric, so it gives different responses to voltages of opposite signs.
Specifically, the pore's conductance is much larger at one sign of the voltage than at the opposite sign.\cite{daiguji_nl_2005,constantin_pre_2007,vlassiouk_nl_2007,karnik_nl_2007,vlassiouk_acsnanno_2008,kalman_am_2008,nguyen_nt_2010}
For this pore, therefore, an important device function emerges: rectification, which is defined as the ratio of the currents in the forward- and reverse-biased states (denoted by ON and OFF states in this work). 

While there are a lot of charge inversion studies in the literature both for planar double layers~\cite{valisko_jpcc_2007,colla_jcp_2016,valisko_aipadv_2018,mashayak_jcp_2018,voukadinova_jcp_2019} and negatively charged pores,~\cite{he_jacs_2009,li_nanoletters_2015,li_jpcc_2019,lin_jacs_2020} studies for the behavior of multivalent ions in bipolar pores are virtually absent.
One notable example is the paper of Li et al.,~\cite{li_jpcc_2019} where the bipolar charge pattern is created by reversible adsorption of multivalent ions on the wall of a symmetric hourglass nanopore only on one side.

In our previous work~\cite{fertig_jpcc_2019} a scaling behavior was studied for a bipolar nanopore using multivalent electrolytes. 
By scaling we mean that the device function, rectification in this case, is a smooth and monotonic function of a universal scaling parameter, $\xi=R^{\mathrm{pore}}/(\lambda\sqrt{z_{+}|z_{-}|})$, where $R^{\mathrm{pore}}$ is the pore radius, $\lambda$ is the electrostatic characteristic screening length of the electrolyte, and $z_{+}$ and $z_{-}$ are the valences of the cation and the anion, respectively.
This result makes it possible to design nanopores of specific rectification properties for various combinations of pore radius and electrolyte concentration, $c$.
We can get the same rectification for small pore radius with large concentration as for large pore radius with small concentration.

The basic explanation is that the ionic current is controlled by depletion zones of the coions in the various regions of the pore.
These depletion zones, on the other hand, appear if the double layers formed near the pore wall in the radial dimension overlap in the pore's centerline.
The degree of overlap depends on the relation of $R^{\mathrm{pore}}$ and $\lambda$.
This scaling works for multivalent electrolytes too (2:2, 2:1, 3:1) if the $\sqrt{z_{+}|z_{-}|}$ factor is included in the $\xi$ scaling parameter.

That work~\cite{fertig_jpcc_2019} was performed for various combinations of $R^{\mathrm{pore}}$, $c$, and $z_{i}$ at fixed surface charge $\sigma=1$ $e$/nm$^{2}$ (the charge pattern was then $\pm \sigma$).
Here, we fix pore radius and salt concentration and change surface charge for various $z_{+}:z_{-}$ electrolytes. 
While our focus was on scaling in our previous work,~\cite{fertig_jpcc_2019} here we concentrate on the $\sigma$-dependence and the anomalous response of the nanopore to the presence of multivalent ions.

By ``anomalous'' we mean both anomalous compared to the 1:1 case, and anomalous compared to mean-field calculations based on the Poisson-Nernst-Planck (PNP) theory.
PNP theory and its variants supplemented with hydrodynamic equations are commonly used in nanopore modeling.~\cite{Ali_ACSnano_2012,tajparast_bba_2015,ramirez_jmembsci_2018,Ali_AC_2018,ramirez_ea_2019,lin_jacs_2020}
It is, however, also a common knowledge that the PNP theory, and its equilibrium limit, the Poisson-Boltzmann (PB) theory, cannot handle many-particle correlations that are beyond the mean field approximation.~\cite{boda_jcp_2002,henderson-pccp-11-3822-2009,voukadinova_jcp_2019}
Therefore, PNP cannot reproduce phenomena that are results of strong ionic correlations, charge inversion, for example.

In this work, we use a particle simulation method, the Local Equilibrium Monte Carlo (LEMC) technique~\cite{boda_jctc_2012} that naturally includes these correlations.
LEMC is an adaptation of the Grand Canonical Monte Carlo (GCMC) method for a non-equilibrium situation.
Coupled to the Nernst-Planck (NP) transport equation, we obtain a method (NP+LEMC) similar to PNP, but the statistical mechanical part (PB) replaced with LEMC.
By comparing the results given by PNP and NP+LEMC we can conclude how accurate PNP is in systems containing multivalent electrolytes.~\cite{fertig_jpcc_2019}
With the wide usage of PNP in calculations for nanopores, this is an important comparison.


\section{Model and applied methods}


\subsection*{Model of a nanopore}
\label{subsec:nanoporemodel}

\begin{figure}[t]
 \centering
\includegraphics[width=0.4\textwidth]{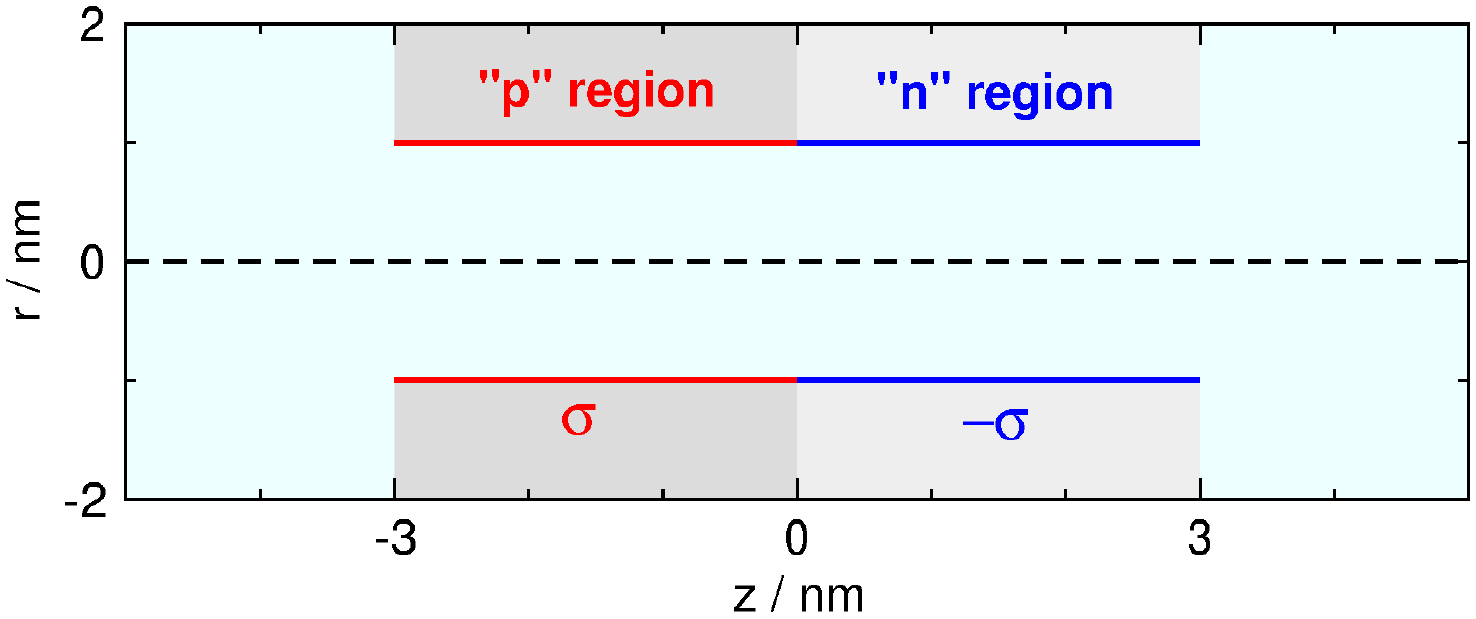}
\caption{The pore used in our simulations} 
 \label{geometry}
\end{figure}

The model of the device was constructed of two baths separated by a membrane.
These two baths are connected with a cylindrical pore through the membrane.
The LEMC simulations are done in the three-dimensional space.
The results presented in this paper will be in cylindrical coordinates due to the rotational symmetry of the system around the axis of the pore.
The simulation domain is a cylinder with a $30$ nm length and $18$ nm radius, while the pore's length is $6$ nm ($H$).
The membrane and the pore are confined by hard walls with a surface charge on the wall of the pore (Fig.\ \ref{geometry}).  

The pore has two charged regions: a positively charged region with surface charge $\sigma$ and a negatively charged region with surface charge $-\sigma$.
The surface charge on the cylinder is modeled as fractional point charges on a $0.2\times 0.2$ nm grid in LEMC, while on a much finer triangular mesh in PNP.
Both of the charged regions have a length of $3$ nm.

\subsection*{Model of the electrolyte and the transport}
\label{subsec:electrolytemodel}

The reduced model applied in this study is based on several approximations.
Interaction of ions with the solvent molecules and the dynamical properties of ions are hidden/coarse-grained in the system's response functions.
Ions are represented as charged hard spheres and the water is modeled as an implicit continuum background.
Screening of the ions' electric field by water molecules is taken into account by one response function, the dielectric constant, $\epsilon$.
Then, the interaction potential between charged hard spheres inside a dielectric is
\begin{equation}
u_{ij}(r)=
\begin{cases}
\infty & \text{if} \quad r < R_i + R_j\\
\dfrac{1}{4\pi\epsilon_0\epsilon}\dfrac{z_i z_j e^{2}}{r} & \text{if} \quad r \geq R_i + R_j
\end{cases}
\end{equation}
where $R_i$ and $z_i$ are radius and the valence of ionic species $i$,  $\epsilon_0$ is the permittivity of vacuum, $e$ is the elementary charge, and $r$ is the distance between the two ions ($R_{i}=0.15$ nm for both ions).
The friction of the ions with the surrounding water molecules is taken account by the other response function, the diffusion coefficient, $D_i(\mathbf{r})$, which is used in the NP transport equation for the ionic flux:
\begin{equation}
\mathbf{j}_i(\mathbf{r})=-\dfrac{1}{k_\mathrm{B}T}D_i(\mathbf{r})c_i(\mathbf{r})\nabla \mu_i(\mathbf{r}),
\label{eq:np}
\end{equation} 
where $\mathbf{j}_i(\mathbf{r})$ is the ionic flux, $c_i(\mathbf{r})$ is the concentration, and $\mu_i(\mathbf{r})$ is the electrochemical potential of ionic species $i$, $k_\mathrm{B}$ is Boltzmann's constant, and $T = 298.15$ K is the temperature.
For the diffusion coefficient profile, $D_{i}(\mathbf{r})$, we use a piecewise constant function, where the value in the baths is $1.334\times 10^{-9}$ m$^{2}$s$^{-1}$ for both ionic species, while it is the tenth of that inside the pore, $D_{i}^{\mathrm{pore}}$, as in our earlier works.~\cite{matejczyk_jcp_2017,madai_jcp_2017,madai_pccp_2018,fertig_jpcc_2019}
These particular choices do not qualitatively affect our conclusions.

To solve the NP-equation a relation should be established between the electrochemical potential profile and the concentration profile. 
This is provided by statistical mechanics. 
Two different approaches are examined in this study: one of them uses a particle simulation method (LEMC), while the other uses the PB theory.
When these methods are coupled to the NP equation, the resulting techniques are called NP+LEMC and PNP, respectively. 
Both methods are based on a self-consistent, convergent iterative procedure in which the results satisfy the continuity equations $\nabla\cdot \mathbf{j}_{i}(\mathbf{r})=0$.

\subsection*{Local Equilibrium Monte Carlo}
\label{subsec:lemc}

The LEMC method, developed by Boda and Gillespie,~\cite{boda_jctc_2012} is an adaptation of the GCMC technique for a non-equilibrium situation, in which $\mu_{i}(\mathbf{r})$ is not constant, so there is no global equilibrium.
There is, however, local equilibrium assumed in small subvolumes defined in the simulational domain denoted by superscript $\alpha$.
Every subvolume is assumed to be in equilibrium locally, so the chemical potential ($\mu_i^{\alpha}$) is constant in each of them.
The set of the $\mu_i^{\alpha}$ values is the input of the LEMC simulations.

In LEMC ion insertion/deletion steps are applied. 
The acceptance probabilities of these are similar to those used in GCMC simulations, except that the local electrochemical potentials, $\mu_i^{\alpha}$, and the volume of the elementary subvolume, $V^{\alpha}$, are used in the formula.
Particle displacement steps can also be used with the usual acceptance probability supplemented with $\mu^{\beta}_{i}-\mu_{i}^{\alpha}$ if the ion is moved from subvolume $\alpha$ to subvolume $\beta$.

The number of ions, therefore, fluctuates in every subvolume.
The output of the simulation is the average number of ionic species $i$ in the volume elements, from which the concentration profile, $c_{i}^{\alpha}$, follows.
Then, the NP equation is used to calculate current densities. 
In an iterative procedure, the $\mu_{i}^{\alpha}$ values are updated until the flux computed from $\mu_{i}^{\alpha}$ data and the resulting $c_{i}^{\alpha}$ data satisfy the continuity equation.
The final result is obtained as running averages.
Further details can be found in the original papers.~\cite{boda_jctc_2012,boda_jml_2014}
 
LEMC has successfully described transport through membranes\cite{boda_jctc_2012,hato_jcp_2012}, calcium channels,~\cite{boda_jml_2014,boda_arcc_2014} bipolar nanopores~\cite{matejczyk_jcp_2017,hato_pccp_2017,fertig_hjic_2017,valisko_jcp_2019}, transistors~\cite{madai_pccp_2018,fertig_mp_2018}, and sensors.~\cite{madai_jcp_2017,madai_jml_2019,madai_pccp_2019}


\subsection*{Poisson-Nernst-Planck theory}
\label{subsec:pnp}

In the PNP theory, ions are not present explicitly in the simulation but rather treated as a continuous function characterizing the local concentration of an ionic species.
The method relies on the solution of the Poisson equation in a domain, which relates the concentration profiles to the mean electrical potential electrostatically.
The solution domain does not include the membrane.
At the same time, the PB theory relates the concentration profiles to the mean electrical potential profile from a statistical mechanical point of view.
Although rarely stated, local equilibrium is assumed and the resulting solution is substituted into the NP equation and iterated until the continuity equation is satisfied.

In PNP, Neumann boundary conditions are defined at the nanopore's wall to produce the desired surface charge pattern.
At the boundaries of the simulation cell, Dirichlet-boundary conditions are used just as in the NP+LEMC method.
To solve the two-dimensional PNP system, the Scharfettel-Gummel scheme is used.~\cite{gummel1964self}
A 2D finite element method is implemented with a triangular mesh which is not uniform: the closer to the pore, the denser the mesh becomes to obtain high accuracy.
Details are found in previous works.~\cite{pietschmann_pccp_2013,matejczyk_jcp_2017}


\section{Results}
\label{sec:results}

We start with showing macroscopic quantities that are the measurable properties of the device.
Here, this is the electrical current carried by an ionic species $i$. 
It is computed as the integral of the $z$-component of $\mathbf{j}_{i}$ over the cross section of the pore:
\begin{equation}
I_{i}=-z_{i}e\int_{0}^{R^{\mathrm{pore}}}\mathbf{j}_{i}(z,r)\cdot \mathbf{k} \; 2\pi r \mathrm{d}r ,
\label{eq:Ii}
\end{equation} 
where $\mathbf{k}$ is the unit vector in the $z$ direction.
This current is the same for all $z$ values inside the pore due to conservation of mass.
The negative sign is there to obtain positive current for a positive voltage (the ground of the electrical potential is on the left hand side).
The experimentally measurable quantity is the total current, $I=\sum_{i}I_{i}$.

In our nanopore studies, we always tended to define quantities of unit dimensions put together from current values.~\cite{hato_pccp_2017,matejczyk_jcp_2017,madai_jcp_2017,madai_pccp_2018,madai_pccp_2019,madai_jml_2019,valisko_jcp_2019}
We called these quantities device functions.
In the case of a bipolar nanopore studied here, the trivial device function is rectification that is defined as $I^{\mathrm{ON}}/I^{\mathrm{OFF}}$, where the absolute values of the total currents in the forward- and reverse-biased states are denoted by $I^{\mathrm{ON}}$ and $I^{\mathrm{OFF}}$ respectively (the OFF-state current is negative).

For the ``pn'' bipolar nanopore shown in Fig.\ \ref{geometry}, the ON state is at the positive voltages, while the OFF state is at the negative voltages.
The current--voltage curves are smooth and monotonic (data not shown), so the behavior of the device can be characterized by fixing the voltage at $200$ mV (ON sate) and  $-200$ mV (OFF state).
Therefore, we show results for the absolute values of the individual and total ionic currents for $200$ and $-200$ mV as functions of $\sigma$.
The bulk salt concentration is fixed at $0.1$ M in this study.
The ionic bulk concentrations, therefore, are $c_{+}^{\mathrm{bulk}}=0.1$ M and $c_{-}^{\mathrm{bulk}}=(z_{+}/|z_{-}|)\times 0.1$ M. 

The anomalous phenomena reported in this work are present at other concentrations too.
Salt concentration $0.1$ M and pore radius $R^{\mathrm{pore}}=1$ nm produce a $\xi$ scaling parameter close to $1$.
This means that the double layers overlap in the pore's centerline and depletion zones are formed.
Similar behavior could be observed at larger pore radii and smaller concentrations providing the same $\xi$ parameter as was shown in our scaling study.~\cite{fertig_jpcc_2019}

The next level in the discussion is digging into the concentration profiles.
The computations provide the $c_{i}(z,r)$ profiles.
It is easier, however, to digest cross-section-averaged axial concentration profiles defined as 
\begin{equation}
c_{i}(z) =  \dfrac{1}{(R_{i}^{\mathrm{cell}}(z))^{2}\pi}  \int\limits_{0}^{R_{i}^{\mathrm{cell}}(z)} c_{i}(z,r) 2\pi r \mathrm{d} r ,
\label{eq:cz}
\end{equation}
where $R^{\mathrm{cell}}_{i}(z)$ is the radius of the simulation cell at coordinate $z$ that is accessible to ion centers. 
Inside the pore, for example, it is $R^{\mathrm{cell}}_{i}=R^{\mathrm{pore}}-R_{i}$.

As was shown in our earlier studies,~\cite{hato_pccp_2017,madai_pccp_2018,valisko_jcp_2019} these axial profiles determine device behavior.
We showed, for example, that if an implicit water model is able to reproduce the qualitative behavior of the axial profiles in comparison with MD simulations with explicit water, then it can reproduce device behavior as well.~\cite{hato_pccp_2017,valisko_jcp_2019}

This statement can be quantified in the slope-conductance approach.~\cite{gillespie_bj_amfe_2008,boda_jgp_2009}
Let us assume that the chemical potential is constant in the radial dimension ($\mu_{i}(z,r)\approx \mu_{i}(z)$) inside the pore.
By substituting the flux density from the NP equation (Eq.\ \ref{eq:np}) into Eq.\ \ref{eq:Ii} we obtain that
\begin{eqnarray}
 I_{i} & \approx & -\dfrac{z_{i}eD_{i}^{\mathrm{pore}}}{kT}  \dfrac{\mathrm{d}\mu_{i}(z)}{\mathrm{d}z} \int\limits_{0}^{R^{\mathrm{pore}}} c_{i}(z,r) 2\pi r \mathrm{d} r \nonumber \\
 & = & -\dfrac{z_{i}eD_{i}^{\mathrm{pore}}}{kT} \dfrac{\mathrm{d}\mu_{i}(z)}{\mathrm{d}z} A_{i} c_{i}(z) ,
 \label{eq:I-cz}
\end{eqnarray} 
where $A_{i}=(R^{\mathrm{pore}}-R_{i})^{2}\pi$ is the effective cross section and Eq.\ \ref{eq:cz} was used.
The principal factor that defines the magnitude of the current is the cross-sectionally averaged axial concentration profile, $c_{i}(z)$, because it is determined by local molecular interactions.
The low-concentration segments of the pore (depletion zones for a given ionic species) can be considered as large-resistance elements inside the pore, and the consecutive segments can be imagined as resistors connected in series.
If a segment has a large resistance, then the whole pore has a large resistance.

Via integration of Eq.\ \ref{eq:I-cz}, we can express the resistance (reciprocal of conductance, $g_{i}$) of the pore for ionic species $i$ as 
\begin{equation}
 g_{i}^{-1} = \dfrac{U}{I_{i}} = -\dfrac{kT}{z_{i}^{2}e^{2}A_{i}D_{i}^{\mathrm{pore}} } \int\limits_{-H/2}^{H/2} \dfrac{\mathrm{d}z}{c_{i}(z)} ,
 \label{eq:resistance}
\end{equation}
where $U$ is the potential drop across the pore.
The resistance, therefore, is associated with the integral of $c_{i}^{-1}(z)$ along the pore.
If $c_{i}(z)$ is very small, $c_{i}^{-1}(z)$ is very large as its integral.
We will show $c_{i}^{-1}(z)$ profiles and resistance values later in this paper (Sections \ref{subsec:off} and \ref{subsec:slopeconductance}).

The rectification is large if $I^{\mathrm{OFF}}$ is small.
$I^{\mathrm{OFF}}$ is small if depletion zones are formed for the ions in the OFF state.
Depletion zones, in turn, are formed if the double layers overlap inside the pore in the radial direction, so the coion is excluded from the pore in these segments. 
Therefore, the radial concentration profiles defined as
\begin{equation}
 c_{i}(r) = \dfrac{1}{H_{2}-H_{1}}  \int\limits_{H_{1}}^{H_{2}} c_{i}(z,r)dz
 \label{eq:cr}
\end{equation} 
must also be studied if we want to understand the behavior of the axial profiles over a given $[H_{1},H_{2}]$ interval.

We will study electrolytes beyond the usual 1:1 electrolyte often used in experiments.
The behavior of 1:1 electrolytes can be reproduced with PNP pretty well.
In this study, one of our goals is to show that PNP fails for electrolytes where multivalent ions are present.

\begin{figure*}[t!]
 \centering
\includegraphics*[width=0.95\textwidth]{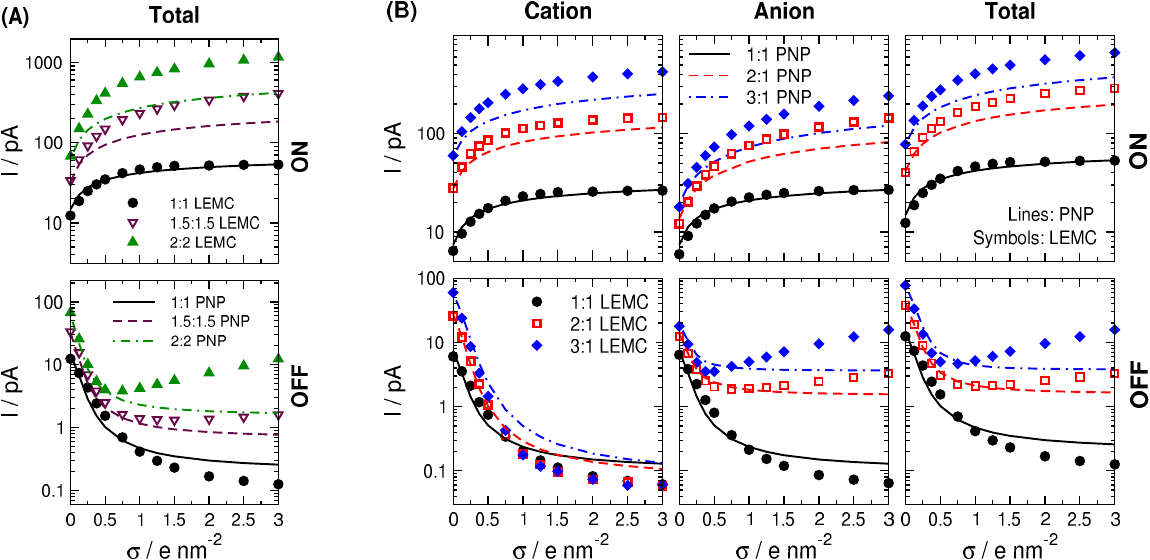}
\caption{Absolute values of ionic currents as functions of surface charge, $\sigma$. (A) Symmetric electrolytes (1:1, 1.5:1.5, and 2:2). Only the total current, $I=I_{+}+I_{-}$, is shown. The individual ionic currents are halves of the total: $I_{+}=I_{-}=I/2$. (B) Asymmetric electrolytes (1:1, 2:1, and 3:1). Currents carried by cations, anions, and the sum of these are shown in the columns from left to right. Top and bottom rows refer to the ON and OFF states, respectively. Symbols and lines refer NP+LEMC and PNP results, respectively, as indicated in the legends.}
 \label{fig02:current}
\end{figure*}

We change ionic charges in two ways.
First, we increase the charges of both cations and anions by simulating a 2:2 electrolyte, and, to follow the trend, the artificial case in between, the 1.5:1.5 system, where $z_{+}=1.5$ and $z_{-}=-1.5$. 
In these cases, increased ionic correlations are responsible for peculiar behavior beyond the 1:1 system, but the electrolyte is still symmetric.

In the other route, we change the valence of cations to $z_{+}=2$ and $3$, for asymmetrical 2:1 and 3:1 electrolytes.
In this case, it is not only the strength of ionic correlations that is responsible for peculiar behavior (such as charge inversion), but also the asymmetry in ionic correlations.

\begin{figure}[b!]
 \centering
\includegraphics*[width=0.25\textwidth]{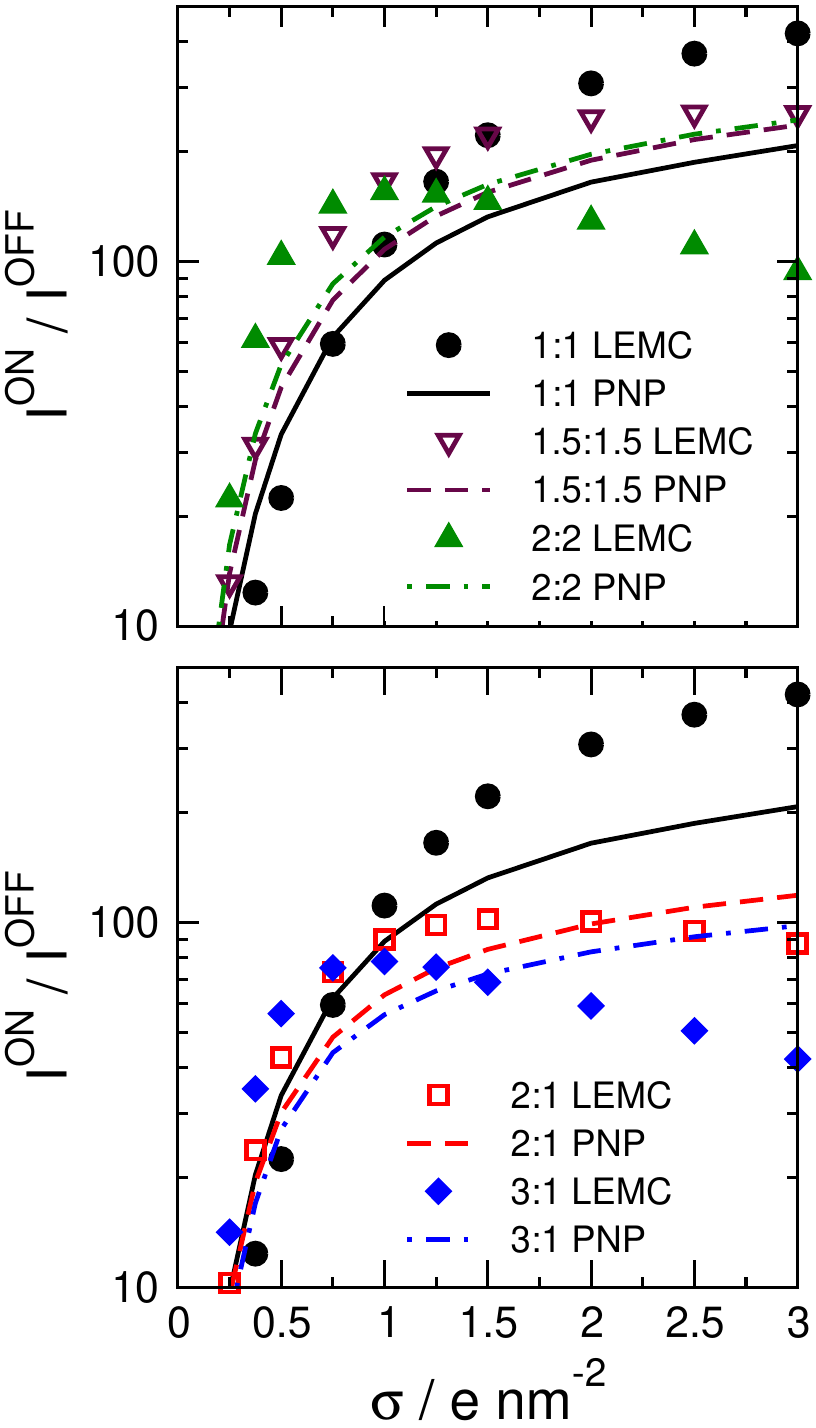}
\caption{Rectification computed from the absolute values of the total currents as $I^{\mathrm{ON}}/I^{\mathrm{OFF}}$ as a function of surface charge, $\sigma$. Top and bottom panels refer to symmetric and asymmetric electrolytes, respectively. Symbols and lines refer NP+LEMC and PNP results, respectively, as indicated in the legends.}
 \label{fig03:rectification}
\end{figure}

Figure \ref{fig02:current} shows the individual and total currents as functions of the surface charge for the ON case (top rows) and the OFF case (bottom rows).
Figure \ref{fig02:current}A shows the results for symmetric electrolytes (1:1, 1.5:1.5, and 2:2), while Fig.\ \ref{fig02:current}B shows the results for asymmetric electrolytes (1:1, 2:1, and 3:1).
The rectification computed from the current data of Figs.\ \ref{fig02:current}A and \ref{fig02:current}B are shown in Fig.\ \ref{fig03:rectification}.
Note that very large, experimentally probably unattainable surface charges are considered (up to $3$ $e$/nm$^{2}$) to make the point.

 \begin{figure*}[t]
 \begin{minipage}[c]{0.75\textwidth}
 \centering
\includegraphics*[width=0.95\textwidth]{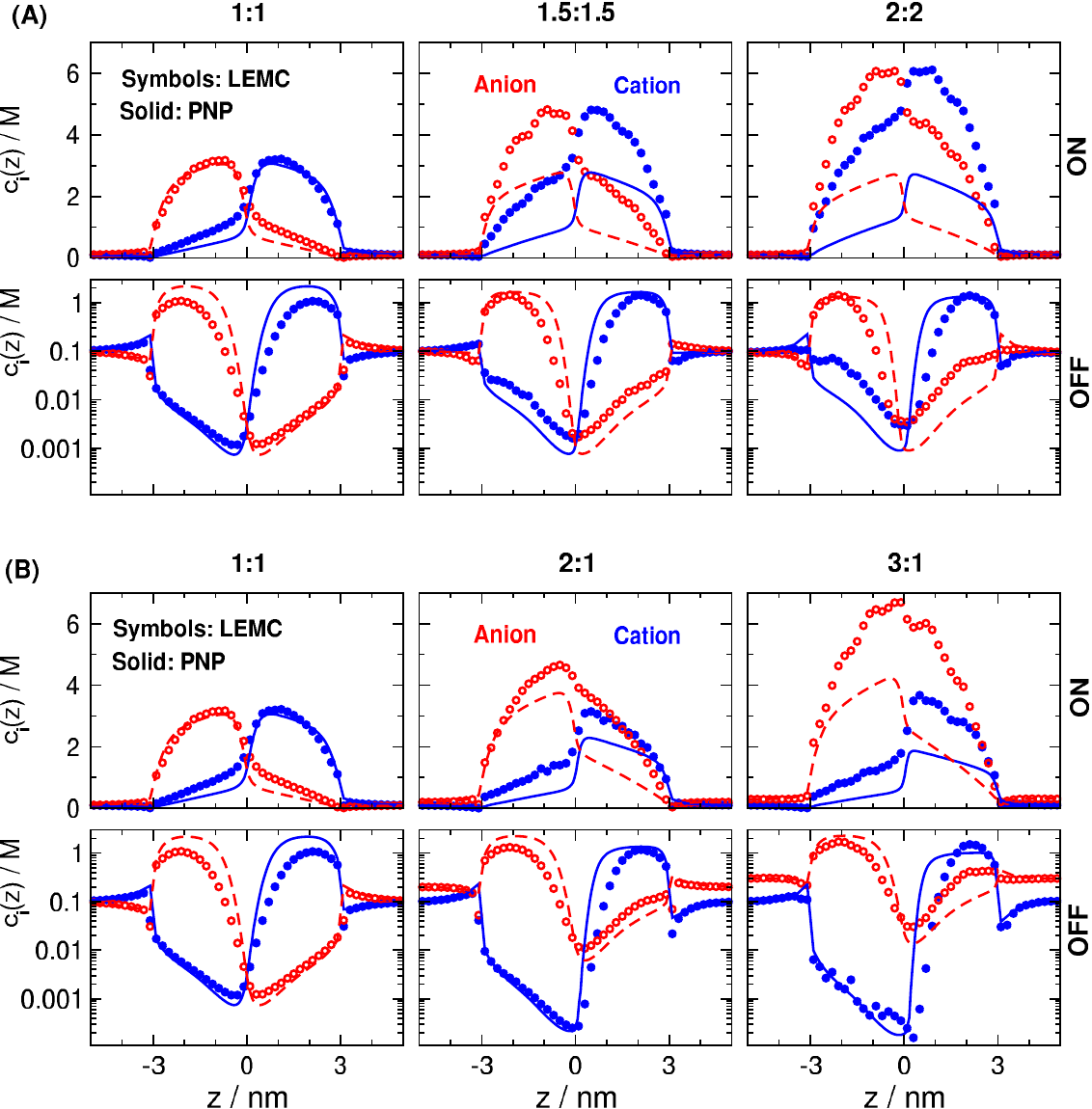}
\end{minipage}\hfill
\begin{minipage}[c]{0.25\textwidth}
\caption{Axial concentration profiles (Eq.\ \ref{eq:cz}) shown for $\sigma=1$ \textit{e}/nm$^{2}$. (A) Symmetric electrolytes (1:1, 1.5:1.5, and 2:2 from left to right). (B) Asymmetric electrolytes (1:1, 2:1, and 3:1 from left to right). Top and bottom rows refer to the ON and OFF states, respectively. Symbols and lines refer to NP+LEMC and PNP results, respectively. Blue and red colors refer to cations, and anions, respectively. Cations are shown with full symbols and solid lines, while anions are shown with open symbols and dashed lines. }
 \label{fig04:axial_sig1}
\end{minipage}
\end{figure*}


\subsection{The ON state}
\label{subsec:on}

In the ON state, currents increase with increasing $\sigma$ because a larger surface charge attracts more counterions into the pore (top rows of Figs.\ \ref{fig02:current}A and \ref{fig02:current}B).
In the OFF state, the absolute values of the currents decrease with increasing $\sigma$ (except when they exhibit a minimum) because depletion zones dominate in the OFF state and larger surface charge excludes more coions from the pore (bottom rows of Figs.\ \ref{fig02:current}A and \ref{fig02:current}B).
As a result, rectification increases with increasing $\sigma$ (except when they exhibit a maximum), which is not a surprise because larger $\sigma$ means a bipolar pore with stronger charge asymmetry (Fig.\ \ref{fig03:rectification}).

What is more interesting about the ON-state results is the change in behavior as ionic valences increase.
As ionic charges get larger, the ON-state currents get larger. 
This is primarily caused by the stronger attraction between the surface charge and the counterion in a given (``p'' or ``n'') region.
This is true both for the NP+LEMC and the PNP results.
While the two methods reproduce the same trend, there is a large quantitative difference between the currents provided by NP+LEMC and PNP.
The NP+LEMC currents are much larger then PNP currents (note the logarithmic scale).

This is the result of stronger ionic correlations and increased concentrations in the pore. 
When cations and anions correlate strongly, the counterions drag the coions with them into the pore even though the surface charge repels the coions. 
This is clearly seen in the top rows of Fig.\ \ref{fig04:axial_sig1}A (symmetric electrolytes) and \ref{fig04:axial_sig1}B (asymmetric electrolytes).
These figures show the axial concentration profiles (Eq.\ \ref{eq:cz}) for a fixed surface charge ($\sigma =1$ \textit{e}/nm$^{2}$).

The cation and anion concentrations are both elevated together with respect to the PNP solution going from left to right (increasing ionic charge).
The difference between the cation and anion profiles is similar in NP+LEMC and PNP because that difference is related to the surface charge.
The magnitudes of the concentrations, however, are increased.
The surface charge attracts counterions into the pore; the counterions attract more coions into the pore; more coions attract more counterions into the pore, and so on.~\cite{voukadinova_jcp_2019}

\begin{figure*}[t]
 \centering
\includegraphics*[width=0.88\textwidth]{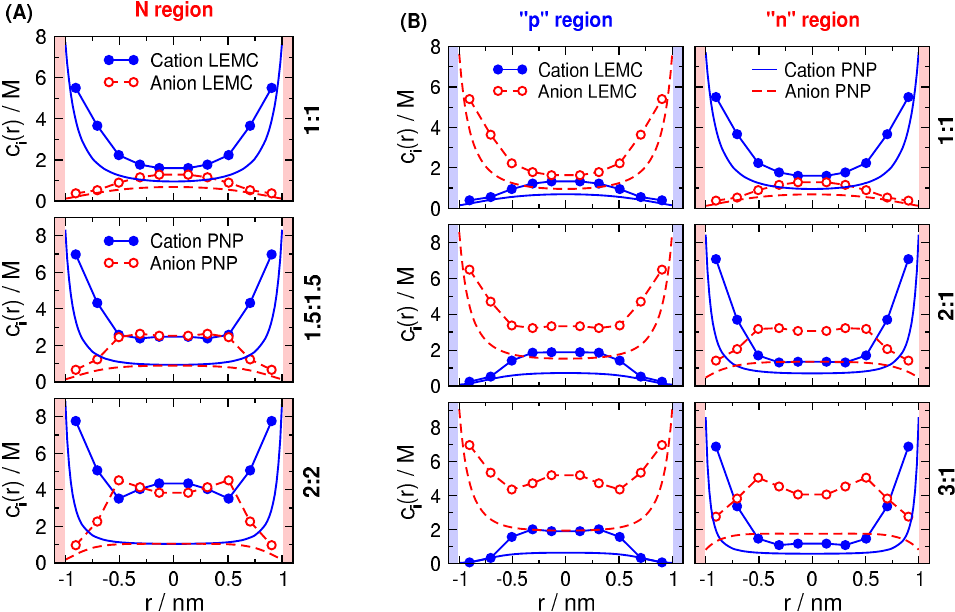}
\caption{The ON-state radial concentration profiles averaged over the respective half regions of the pore (Eq.\ \ref{eq:cr}) for different electrolytes at fixed $\sigma=1$ \textit{e}/nm$^{2}$. (A) For the symmetric electrolytes, only the ``n'' region is shown; the profiles in the ``p'' region look the same with the colors exchanged. (B) For the asymmetric electrolytes, both ``p'' and ``n'' regions are shown because they behave differently due to ion charge asymmetry. Cation valences increase from top to bottom. Symbols and lines refer to NP+LEMC and PNP results, respectively. Blue and red colors refer to cations and anions, respectively. Cations are shown with full symbols and solid lines, while anions are shown with open symbols and dashed lines. Surface charges of regions are shown by vertical lines light blue and red for the ``p'' and ``n'' regions, respectively. The radial profiles are mirrored with respect to $r=0$ and also shown for negative $r$ values for better visualization.}
 \label{fig05:radial_on_sig1}
\end{figure*}

This phenomenon is well visible in Fig.\ \ref{fig05:radial_on_sig1}, which shows radial concentration profiles (Eq.\ \ref{eq:cr}).
Figure \ref{fig05:radial_on_sig1}A shows the profiles for the symmetric case (ON state, $\sigma=1$ \textit{e}/nm$^{2}$) in the ``n'' region (the ``p'' region looks similar to the roles of the ions exchanged).
Not only the concentration on the centerline is larger than the bulk concentration ($0.1$ M) in this confined system but it increases further as $z_{+}=|z_{-}|$ increases. 
This is the reason for the increased axial concentration in Fig.\ \ref{fig04:axial_sig1}A. 

In the 2:2 case, even charge inversion can be observed, a phenomenon characteristic of strong electrostatic ionic correlations. 
In the ON state, however, charge inversion has secondary importance.
It is the elevated concentration due to stronger ionic correlations that is of primary importance.
We will discuss charge inversion in more detail for the asymmetric electrolytes.

The asymmetric case is more complicated because cations and anions behave differently.
Looking at Fig.\ \ref{fig02:current}B, one can observe that cation currents are larger than anion currents for all $\sigma$ if $z_{+}>|z_{-}|$ (2:1 and 3:1 cases).
The explanation is not that cations carry more charge, as there are more anions in the electrolyte due to stoichiometry. 
The reason is that the driving force for multivalent cations is larger at a given voltage: the electrochemical potential difference between the two bulks is $z_{+}eU$ (interaction with the average electric field).

When we look at the axial concentration profiles (Fig.\ \ref{fig04:axial_sig1}B, top row), we can observe that the NP+LEMC curves are larger than the PNP curves in the 2:1 system, and, especially, in the 3:1 system.
This is true for the cations, but even more so for the anions.

This behavior can be better understood by examining the radial profiles (Fig.\ \ref{fig05:radial_on_sig1}B).
Here, we need to consider the ``p'' and ``n'' regions separately because they behave differently due to ionic charge asymmetry.
In every case, there is an excess of counterions at the pore wall ($|r|\sim 1$ nm), but in the asymmetric cases (2:1 and 3:1) there are two phenomena that we already observed in the symmetric case: (1) ionic concentrations are increased in the pore center, and, in the meantime, (2) the concentration of the coion is disproportionately increased.
This latter phenomenon is charge inversion.

This was already observed in the 2:2 case. 
The surface charge attracted the counterions strongly and repelled the coions strongly.
This leads to an overcharge of the wall and an excess of coions in the second layer.

Charge inversion also appears in the asymmetric case, as is well known from numerous studies in the last decades.~\cite{valisko_jpcc_2007,he_jacs_2009,nasir_jcis_2019,ramirez_jms_2018,voukadinova_jcp_2019,li_nanoletters_2015,ramirez_ea_2019}
The emphasis, however, was on the case of a negatively charged wall and multivalent cations (the ``n'' region, in this study).
It is well known that the negative surface charge attracts the multivalent cations strongly resulting in an overcharge and in a charge inversion in the second layer.

We observe in Fig.\ \ref{fig05:radial_on_sig1}B, however, that charge inversion also appears in the ``p'' region, where the monovalent anions are the counterions. 
The mechanism of overcharging, in this case, is not that the wall attracts the counterions too strongly, but that it repels the coions too strongly.
The positive surface charge repels the trivalent cations very strongly that leads to an overcharge of the wall by the anions.

The energetics of these mechanisms can be studied via well-defined terms of the excess chemical potential as shown in earlier papers for the electrical double layer~\cite{valisko_jpcc_2007,voukadinova_jcp_2019} and ion channels.~\cite{gillespie_bj_2008_energetics,boda_jcp_2011_analyze,boda_jcp_2013,boda_cmp_2015}
We defer the energetic studies for the case of nanopores to future publications.

\begin{figure}[b!]
 \centering
 \includegraphics*[width=0.45\textwidth]{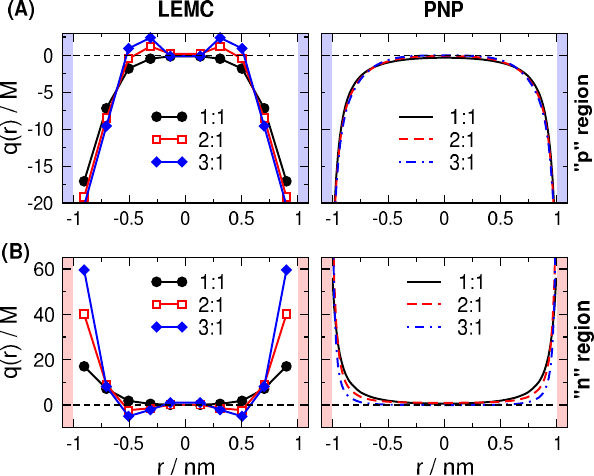}
\caption{ON-state radial charge profiles computed as $q(r)=z_{+}c_{+}(r)+z_{-}c_{-}(r)$ shown for (A) the ``p'' region and (B) the ``n'' region for $\sigma=1$ \textit{e}/nm$^{2}$. The asymmetric electrolytes (1:1, 2:1, and 3:1) are shown with different colors as indicated in the legends. Symbols and lines refer to NP+LEMC (left panels) and PNP (right panels) results, respectively. The radial profiles are mirrored with respect to $r=0$ and also shown for negative $r$ values for better visualization.}
 \label{fig06:radial_charge}
\end{figure}

The charge inversion effects are more readily seen in Fig.\ \ref{fig06:radial_charge}, which shows the radial charge profiles defined as $q(r)=z_{+}c_{+}(r)+z_{-}c_{-}(r)$ (the unit is M).
Also, this figure clearly shows the absence of charge inversion in PNP.
Charge inversion is a key phenomenon that highlights the shortcomings of the mean field treatment of PB-type theories.~\cite{boda_jcp_2002,voukadinova_jcp_2019,mashayak_jcp_2018,ramirez_jmembsci_2018,perezmitta_cpc_2016,fuest_analchem_2017,wang_jpcc_2018}

To summarize, in the ON state, the electrolyte has a high density in the pore, so the dominant effect that originates from larger ionic charges is that cations and anions correlate more strongly. 
That elevates the concentrations of both ionic species inside the pore that, in turn, results in increased currents.
The ion--surface charge correlations are also stronger, contributing to charge inversion, but the larger currents  in excess to the PNP current are mainly determined by the electrostatic ion-ion correlations beyond mean field. 

In the case of the OFF state ($-200$ mV), as we will see in the following subsection, the situation is more complicated.


\subsection{The OFF state}
\label{subsec:off}

As with the ON state, let us start with examining the currents (Fig.\ \ref{fig02:current}).
The OFF currents of individual ions as functions of $\sigma$ first decline, then in some cases they keep declining (monotonic behavior), while in other cases they go through a minimum and start increasing.
For the 1:1 case, the cations and anions have monotonic behavior as produced by both methods.
Increasing $\sigma$ produces deeper depletion zones for cations (anions) in the ``p'' region (``n'' region) that, in turn, result in smaller currents.
Ionic correlations that would counterbalance the dominant surface charge vs.\ coion repulsion are weak in the 1:1 case.
The axial profiles (left panels of Fig.\ \ref{fig04:axial_sig1}) show the formation of these depletion zones (again, note the logarithmic scale).
These profiles also show the good agreement between NP+LEMC and PNP results for the 1:1 system.

When ionic charges are increased (either cation and anion charges simultaneously or cation charges only), nonmonotonic behavior appears in the NP+LEMC results (PNP never reproduces this behavior).
In the symmetric cases (1.5:1.5 and 2:2) both cation and anion currents have a minimum (Fig.\ \ref{fig02:current}A), while in the asymmetric cases (2:1 and 3:1) only the anions have a minimum (Fig.\ \ref{fig02:current}B).
In general, an ionic species shows the minimum if the other ionic species is multivalent.

It is easier to explain this behavior by examining the concentration profiles.
Looking at the OFF axial profiles in Fig.\ \ref{fig04:axial_sig1}A for the 1.5:1.5 and 2:2 cases and in Fig.\ \ref{fig04:axial_sig1}B for the 2:1 and 3:1 cases, the profiles given by NP+LEMC and PNP look qualitatively similar.
The multivalent cation shows a very deep depletion zone in the ``p'' region with both methods.
The anomalous behavior is shown by the anions in the ``n'' regions, but this figure does not explain the $\sigma$--dependence because $\sigma$ is fixed ($1$ \textit{e}/nm$^{2}$).

\begin{figure}[t]
 \centering
\includegraphics*[width=0.48\textwidth]{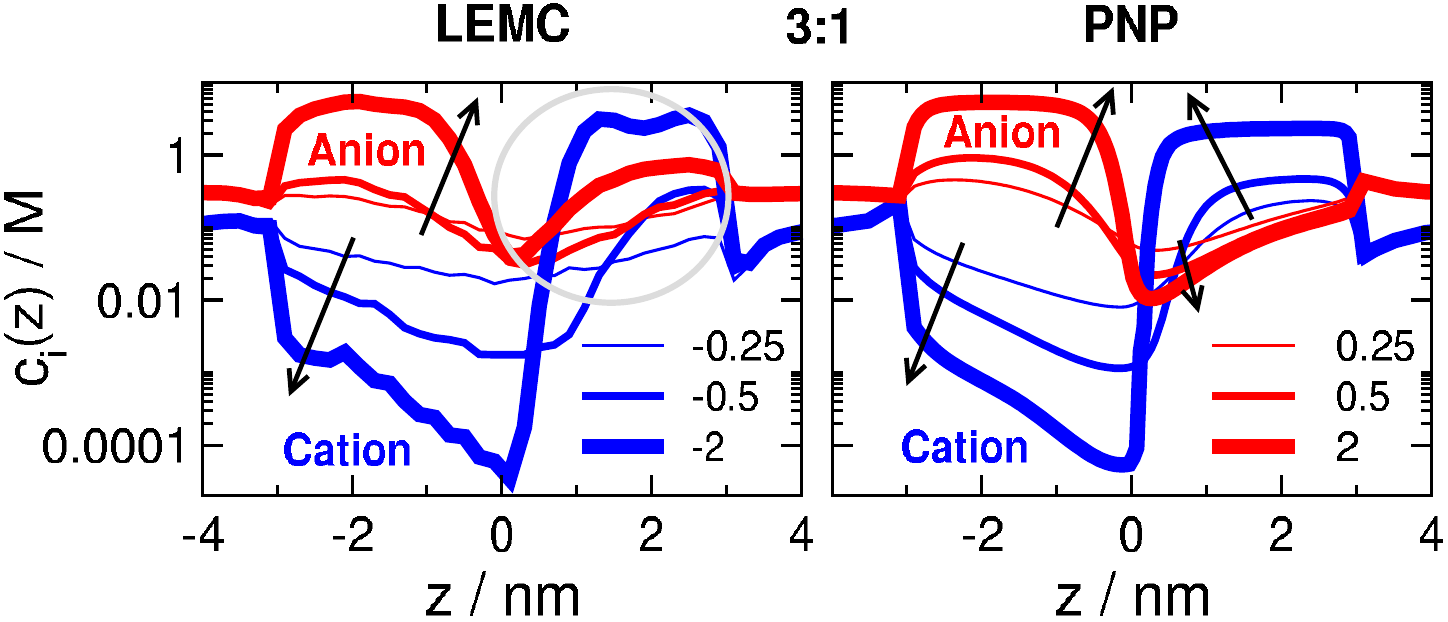}
\caption{OFF-state axial concentration profiles for the 3:1 case at $\sigma=0.25$, $0.5$, and $2$ \textit{e}/nm$^{2}$ (indicated by lines with increasing thickness). Left and right panels refer to NP+LEMC and PNP results, respectively. Blue and red lines refer to cations and anions, respectively. Arrows show the direction of increasing $\sigma$.}
 \label{fig07:axial_off_sigdep_c}
\end{figure}

To understand the $\sigma$--dependence of the axial profiles, we plot them for various $\sigma$ values.
We selected $\sigma=0.25$, $0.5$, and $2$ \textit{e}/nm$^{2}$ because they are representative values below, at, and above the minimum.
We plot the profiles only for the 3:1 case because it is representative of the 2:1 and 2:2 cases as well (Fig.\ \ref{fig07:axial_off_sigdep_c}).

There is no surprise in the cation profiles (blue lines).
Their depletion in the left ``p'' region becomes stronger as $\sigma$ increases (indicated by increasing line thickness and black arrows).
Also, the cation peaks become larger in the ``n'' region with increasing $\sigma$.
Both the ``p''-region and the ``n''-region behaviors are reproduced by PNP.
The agreement between NP+LEMC and PNP is also shown by the current profiles (bottom-left panel of Fig.\ \ref{fig02:current}B).
In these cases, the strong correlation between the cations and the surface charge dominates.

The anion profiles, on the other hand, show a different trend.
At a small surface charge ($\sigma=0.25$ \textit{e}/nm$^{2}$, thin lines), the peaks are not as high, the depletion zones are not as deep, and NP+LEMC and PNP produce similar profiles.
As the surface charge increases ($\sigma=0.5$ \textit{e}/nm$^{2}$, lines of medium thickness), deviations between NP+LEMC and PNP appear, but the trend is the same: the anion concentration increases in the ``p'' region, and decreases in the ``n'' region.

The trend breaks in the ``n'' region (right) when the surface charge increases further ($\sigma=2$ \textit{e}/nm$^{2}$, thick lines).
In PNP, the anion concentration decreases further in the ``n'' region, as expected due to increasing repulsion between the anions and the negative surface charge.
In NP+LEMC, however, the anion concentration increases in the ``n'' region (indicated by the gray circle) due to strong correlations between cations and anions.

\begin{figure}[t!]
 \centering
\includegraphics*[width=0.48\textwidth]{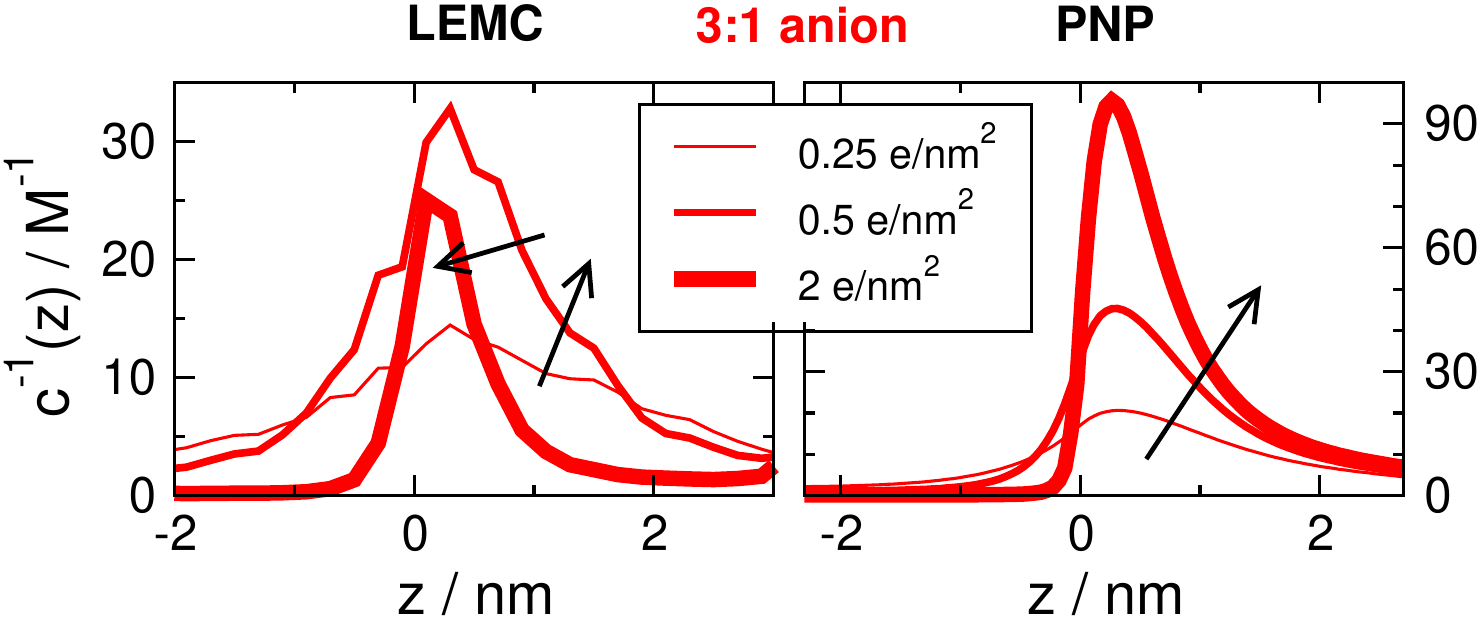}
\caption{Reciprocals of the anion concentration profiles shown in Fig.\ \ref{fig07:axial_off_sigdep_c} for the 3:1 case at $\sigma=0.25$, $0.5$, and $2$ $e/\mathrm{nm}^{2}$ (indicated by lines with increasing thickness). Left and right panels refer to NP+LEMC and PNP results, respectively. Arrows show the direction of increasing $\sigma$.}
 \label{fig08:axial_off_sigdep_cinv}
\end{figure}

How these concentration profiles lead to the nonmonotonic behavior of the anion current in the OFF state (Fig.\ \ref{fig02:current}B) is not readily apparent from the profiles in Fig.\ \ref{fig07:axial_off_sigdep_c}. 
Therefore, we plotted the same anion concentration profiles in Fig.\ \ref{fig08:axial_off_sigdep_cinv}, but now the reciprocal concentration, $c_{-}^{-1}(z)$, instead of $c_{-}(z)$.
Also, the scale for the $c_{-}^{-1}(z)$ profile is linear, while it is logarithmic for the $c_{-}(z)$ profile.
Depletion zones correspond to peaks in the $c_{-}^{-1}(z)$ profile.

Plotting $c_{-}^{-1}(z)$ is advantageous because the area under the curve is proportional to the resistance of the pore (Eq.\ \ref{eq:resistance}).
For PNP, the $c_{-}^{-1}(z)$ profiles increase with increasing $\sigma$ (increasing $\sigma$ is indicated by the arrows), so their integrals also increase, and, consequently, the pore resistance increases. 
The anion current, therefore, decreases monotonically.

\begin{figure*}[t!]
 \centering
\includegraphics*[width=0.85\textwidth]{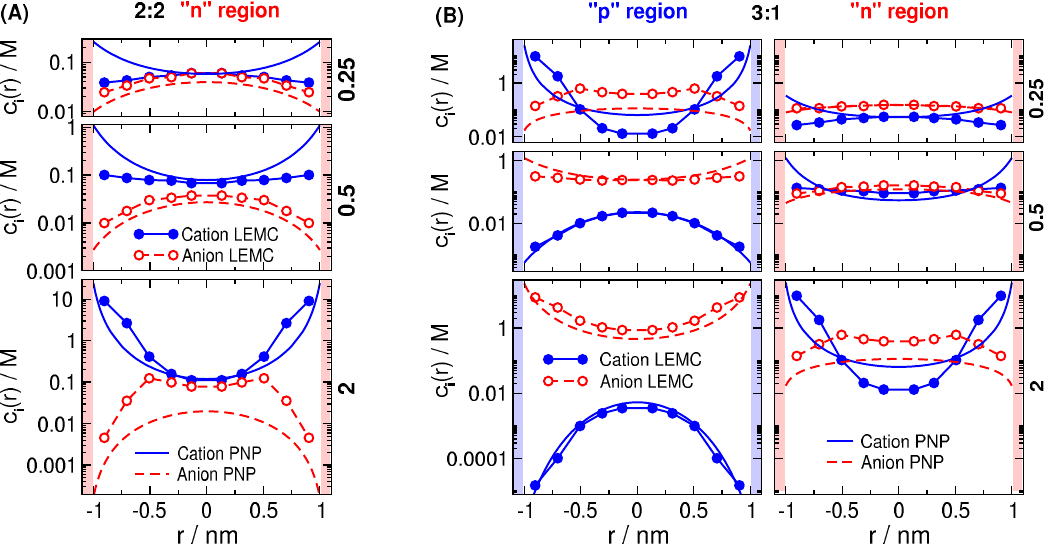}
\caption{OFF-state radial concentration profiles averaged over the respective half regions of the pore (Eq.\ \ref{eq:cr}) for varying surface charges. (A) In the symmetric 2:2 case, only the ``n'' region is shown; the profiles in the ``p'' region look the same with the colors exchanged. (B) In the asymmetric 3:1 case, both ``p'' and ``n'' regions are shown because they behave differently due to ion charge asymmetry. Surface charge increases from top to bottom ($\sigma=0.25$, $0.5$, and $2$ $e/\mathrm{nm}^{2}$). Symbols and lines refer to NP+LEMC and PNP results, respectively. Blue and red colors refer to cations, and anions, respectively. Cations are shown with full symbols and solid lines, while anions are shown with open symbols and dashed lines. Surface charges of regions are shown by vertical lines light blue and red for the ``p'' and ``n'' regions, respectively. The radial profiles are mirrored with respect to $r=0$ and also shown for negative $r$ values for better visualization.}
 \label{fig09:radial_sym_OFF-sigdep}
\end{figure*}

For NP+LEMC, this trend is observed only between $\sigma=0.25$ and $0.5$ \textit{e}/nm$^{2}$ (follow the arrow).
The $\sigma=2$ \textit{e}/nm$^{2}$ curve (thick line) is below the line for $\sigma=0.5$ \textit{e}/nm$^{2}$, so its integral is smaller.
The explanation of this behavior is that ionic concentrations start to be large enough inside the pore at $\sigma=2$ \textit{e}/nm$^{2}$ so that ion-ion correlations begin to contribute.

This is better seen by examining radial profiles.
Figure \ref{fig09:radial_sym_OFF-sigdep} shows radial profiles like in Fig.\ \ref{fig05:radial_on_sig1}, but now the rows refer to different surface charges ($\sigma=0.25$, $0.5$, and $2$ \textit{e}/nm$^{2}$).
Figure \ref{fig09:radial_sym_OFF-sigdep}A shows a symmetric case (2:2) and Fig.\ \ref{fig09:radial_sym_OFF-sigdep}B an asymmetric case (3:1).

Figure \ref{fig09:radial_sym_OFF-sigdep}A shows the ``n'' region, so the anions are the coions (red).
Elevated ionic concentrations can be observed for $2$ \textit{e}/nm$^{2}$ together with a bump in the anion profiles that indicates a charge inversion-like phenomenon (this can be seen better for the 3:1 case later).
Strong correlations between the divalent cations and the $2$ \textit{e}/nm$^{2}$ surface charge increase the cation concentration in the double layer near the wall which, in turn, increases anion concentrations due to strong correlations between ions.

For the 3:1 case, we show both regions (Fig.\ \ref{fig09:radial_sym_OFF-sigdep}B).
Again, the bottom row for $2$ \textit{e}/nm$^{2}$ shows very strong depletion for the trivalent cations in the ``p'' region.
In the ``n'' region, we observe charge inversion, where a strong correlation between cations and the surface charge causes overcharging that, in turn, leads to a charge inversion in the centerline of the pore ($|r|\sim 0$ nm).
This phenomenon, of course, is completely absent in PNP.

To summarize, there are two competing effects that appear in electrolytes beyond the simple 1:1 system, where PNP works fine.
These effects appear when one or two ionic species have multiple valence.
The two effects are (1) the increased correlation between ions and (2) the increased correlations between multivalent ions and surface charge.
In the ON state, the increased correlations between ions is the dominant effect because concentrations are large in the nanopore, so ions are close to each other and they are forced to correlate.

In the OFF state, on the other hand, the balance of the two effects results in anomalous behavior.
Ions are depleted in the pore at small $\sigma$, so neither ion-ion correlations nor ion-wall correlation can cause unusual phenomena.
At large surface charges, however, ion-wall correlations become strong which partly elevates average ion concentrations in the pore and partly causes overcharge. 
Increased ion concentrations cause the ion-ion correlations to work together with ion-wall correlations and cause anomalously increased concentrations (and, thus, currents) and charge inversion in this case.


\subsection{Rectification}
\label{subsec:rectification}

The ratio of the ionic currents shown in Fig.\ \ref{fig02:current} produces the rectification behavior shown in Fig.\ \ref{fig03:rectification}.
Because we divide a monotonic function (ON current) with a nonmonotonic one with a minimum (OFF current), we obtain curves exhibiting maxima in the interesting cases (2:2, 2:1, and 3:1).

Rectification is an important device function investigated in many experimental and modeling studies.~\cite{daiguji_nl_2005,constantin_pre_2007,vlassiouk_nl_2007,karnik_nl_2007,vlassiouk_acsnanno_2008,kalman_am_2008,nguyen_nt_2010}
Because modeling of ion diffusion is usually based on PNP, it is worth studying how accurate is PNP in reproducing the results of LEMC simulations.

For $\sigma$ values, where the NP+LEMC rectification curves are increasing functions of $\sigma$, the difference between the PNP and LEMC results is quantitative.
In this regime, PNP considerably ($1.5-2$ times) underestimates the rectification of NP+LEMC.
This is an important finding because it appears at low surface charges, where ionic correlations are not at a full strength.
The quantitative differences between PNP and LEMC in this $\sigma$ regime are a clear indication of the errors introduced by the mean field treatment of multivalent ions by PNP.

The full power of ionic correlations occurs at high surface charges where they cause not only quantitative, but qualitative deviations from the PNP behavior.
The NP+LEMC rectification curves go through a maximum, a behavior that is absent in PNP.
This behavior is robust, appearing in every electrolyte above 1:1.
The maximum appears at smaller $\sigma$ for electrolytes with stronger ionic correlations.
Note that the strength of ionic correlations can be characterized by the $z_{\mathrm{if}}=\sqrt{z_{+}|z_{-}|}$ variable, as was shown in our previous paper,~\cite{fertig_jpcc_2019} where we proposed a scaling behavior for multivalent electrolytes.
Based on our results, we propose using PNP with extra care when multivalent ions are present.


\subsection{Selectivity}
\label{subsec:selectivity}

\begin{figure}[t!]
 \centering
\includegraphics*[width=0.48\textwidth]{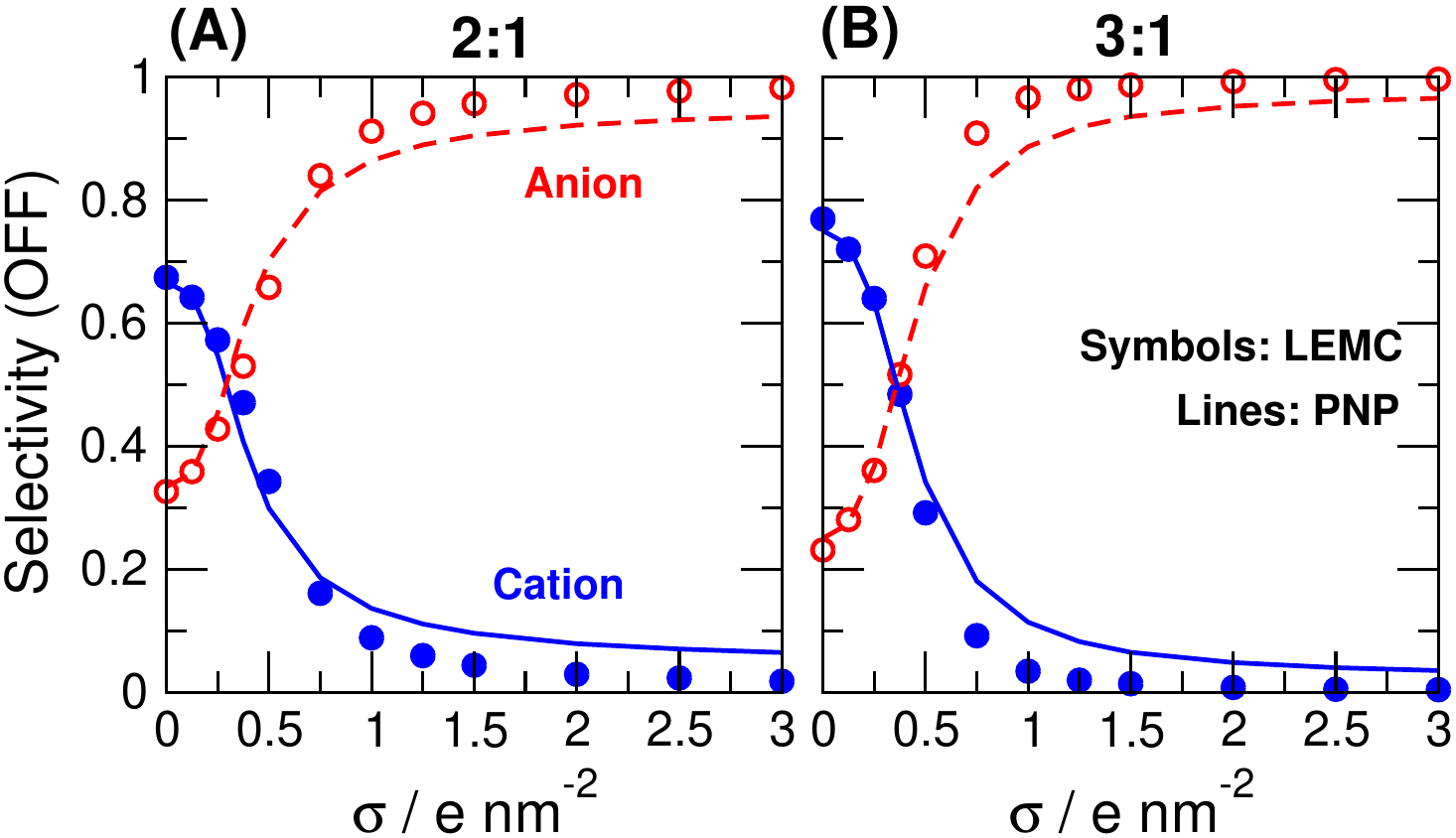}
\caption{Selectivity defined as $I_{i}/I$ for (A) the 2:1 and (B) the 3:1 cases as a function of the surface charge. Symbols and lines refer to NP+LEMC and PNP results, respectively. Blue and red colors refer to cations and anions, respectively.}
 \label{fig10:off_selectivity}
\end{figure}

From the raw data for currents (Fig.\ \ref{fig02:current}) another interesting device functions can be deduced.
This function is selectivity, which was also investigated in many papers dedicated to the behavior of nanopores\cite{hato_pccp_2017,valisko_jcp_2019,fertig_jpcc_2019}. 
Here, we define selectivity as $S_{i}=I_{i}/I$, namely, the current carried by a certain ionic species $i$ divided by the total current.
If the nanopore is non-selective, this value is $1/2$, $2/3$, and $3/4$ for symmetric, 2:1, and, 3:1 electrolytes, respectively.
If this value is close to $1$, the nanopore is selective for ionic species $i$.
Note that defining selectivity in terms of particle currents instead of electrical currents, the non-selective case would be characterized by the value $1/2$ for all electrolytes.

Selectivity is interesting only for asymmetric electrolytes (2:1 and 3:1) because the pore is perfectly non-selective for the symmetric electrolytes.
This follows from both the symmetry of the bipolar nanopore (the ``p'' and ``n'' regions are of equal length and equal charge in magnitude) and the symmetry of the ions ($z_{+}=|z_{-}|$, $R_{+}=R_{-}$, and $D_{+}=D_{-}$).

In the case of asymmetric electrolytes, we obtain the $\sigma$-dependence shown in Fig.\ \ref{fig10:off_selectivity} for the OFF case (the ON case is less interesting showing cation selectivity for all $\sigma$ values).
The PNP and LEMC methods provide similar selectivity behavior despite LEMC having a minimum in the OFF current.
This is because the cation current decreases steeply after the minimum in both methods.
At large $\sigma$, the surface charge repels the multivalent cations more strongly than the monovalent anions.
The depletion zone of the cations, therefore, is deeper than those of the anions.
The current of the cations, as a consequence is smaller than those of the anions for large surface charges.

Charge inversion causes the small difference between the PNP and LEMC selectivity values above $\sigma \approx 0.5 $ $e$/nm$^{2}$.
The larger anion selectivity of LEMC is the result of the anion leakage shown by this method.

Selectivity often works on the basis of attraction when a confined system attracts one species more than another species. 
There is also another mechanism, however, present in poorly occupied pores, that is based on exclusion rather than attraction.
That is, the pore repels one species more than other species.
This was the mechanism, for example, in neural sodium channels that excluded the large K$^{+}$ ions more than the small Na$^{+}$ ions~\cite{boda_pccp_2002,boda_bj_2007,csanyi_bba_2012} based on volume exclusion.
Here the exclusion is electrostatic in origin.


\subsection{Slope conductance}
\label{subsec:slopeconductance}

Finally, we apply the slope conductance analysis that we exploited before in the case of ion channels~\cite{gillespie_bj_amfe_2008,boda_jgp_2009,malasics_bba_2009,malasics_bba_2010,giri_pb_2011} and nanopores~\cite{gillespie_bj_nanopore_2008,valisko_jcp_2019} to relate local ionic concentration to the resistance of a segment of the pore.
The resistance of a $[H_{1},H_{2}]$ segment can be estimated by integrating the $c_{i}^{-1}(z)$ profile (Eq.\ \ref{eq:resistance}).

This is especially advantageous in the case of the bipolar nanopore studied here because the ``p'' and ``n'' regions have characteristically distinct and different conduction properties due to their surface charges of opposite signs.
Current is a property of the whole pore because it measures the average number of ions passing through the pore in a time unit.
By computing the resistances of the ``p'' and ``n'' regions individually we can say something quantitative about the properties of those regions, separately.

We perform the integration for three regions: $[-H/2,-0.5\, \mathrm{nm}]$ (the ``p'' region), $[0.5\, \mathrm{nm}, H/2]$ (the ``n'' region), and $[-0.5\, \mathrm{nm}, 0.5\, \mathrm{nm}]$ (``junction'' region).
The intermediate $1$ nm thick ``junction'' region is handled separately because it has ``mixed'' features, while we want to assess the properties of the ``clean'' ``p'' and ``n'' regions.

\begin{figure}[t!]
 \centering
\includegraphics*[height=0.30\textwidth]{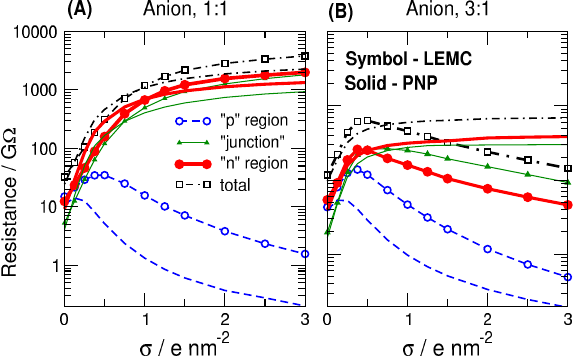}
\caption{Resistances of various regions inside the pore along the $z$ axis, as computed with the slope-conductance method (Eq.\ \ref{eq:resistance}). Panels (A) and (B) refer to the 1:1 and 3:1 cases, respectively. The results are shown for the anions. The regions are the ``p'', ``n'', and a ``junction'' region as defined in the text. Symbols and lines represent LEMC and PNP results.}
 \label{fig11:slope_cond}
\end{figure}

Fig.\ \ref{fig11:slope_cond} shows the results for the anion (the interesting species showing anomaly in the 3:1 case) for the 1:1 and 3:1 electrolytes.
The behavior of the ``p'' region (dashed blue lines) is the same in the NP+LEMC and PNP treatments: the resistance of this region decreases with increasing $\sigma$ because increasing positive surface charge attracts more and more anions into this region.
The increase at small $\sigma$ values is the effect of the adjacent ``n'' region; our definition of the ``junction'' region is not wide enough, it seems.
Practically, this is the result of our relatively short pore, where the neighboring regions ``communicate'' with each other.
We would see an even more ``clear-cut'' behavior in the case of longer pores.

The ``n'' region is the interesting region (thick red lines), where the anions are excluded by the negative surface charge, as seen from the increasing resistance given by PNP.
The ionic correlations between cations and anions, however, bring more anions into this region resulting in a decreasing resistance above $\sigma \sim 0.5$ $e$/nm$^{2}$ as seen from the NP+LEMC data.
The resistance of the ``n'' region dominates the total resistance (note the logarithmic scale), resulting in the nonmonotonic behavior of the anion current.


\section{Closing remarks}
\label{sec:summary}

In this work, we examined the transport of multivalent ions through a bipolar nanopore as a function of the surface charge strength ($\sigma$ and $-\sigma$ in the ``p'' and ``n'' regions were changed simultaneously).
We computed the currents both at the ON and the OFF state ($\pm 200$ mV).
This system proved to be a good test case, because anomalous device behavior could be observed on the basis of electrostatic correlations only.

Increasing $\sigma$ caused stronger correlations between the ions and surface charge.
Note that this correlation is either attractive (in relation of surface charge and counterions) or repulsive (in relation of surface charge and coions).
Both are present in the bipolar nanopore.
Increasing ionic charges (only $z_{+}$ or both valences) caused stronger attractive correlation between cations and anions.

All these stronger correlations produced differences between the mean-field PNP theory and the LEMC simulations, which naturally included all the correlations beyond mean field.
We found different phenomena in the ON and the OFF state.

In the ON state, the applied field favors the presence of ions, so ionic concentrations were elevated.
The main difference between the PNP and the LEMC results was that the concentrations were elevated more in LEMC due to the presence of strong correlations between cations and anions, especially if one or both were multivalent. 
Regarding currents, this caused a quantitative deviation: LEMC provided much larger currents than PNP.

In the OFF state, depletion zones of coions are formed, and control the behavior of the system.
We found nonmonotonic behavior in the current of the ionic species when the other ionic species is multivalent as a function of $\sigma$. 
These are the anion in the 2:1 and 3:1 cases and both ions in the 2:2 case.  
The key phenomenon is the leakage of this ionic species.
By ``leakage'', we mean that the current of this species increases with $\sigma$ (above a threshold) in LEMC, while in PNP it does not. 
The leakage is absent in the 1:1 case.
In the 2:1 and 3:1 systems, we observe anion leakage.
In the 2:2 case, we observe leakage of both species.
The leakage is larger at large $\sigma$ values where the LEMC data show increasing deviation from the PNP data.

Deviations of PNP results from LEMC results indicate that the mean-field PNP theory should be used with extra care when multivalent ions are present.


\section*{Acknowledgements}
\label{sec:ack}

We acknowledge the financial support of the National Research, Development and Innovation Office -- NKFIH K124353. 
Present article was published in the frame of the project GINOP-2.3.2-15-2016-00053.
The advices of Dirk Gillespie are gratefully acknowledged.

\end{document}